\documentclass[english]{eptcs}

\usepackage{prooftree,amsmath,amssymb,wasysym,xspace}
\usepackage{mathptmx,stmaryrd}
\usepackage{eurosym}
\usepackage[T1]{fontenc}
\usepackage[mathscr]{euscript}
\usepackage[usenames,dvipsnames]{color}
\usepackage{epsfig,ccd}
\usepackage{breakurl}

\newcommand{\processes}[1]{\vspace{0.1cm}
\centerline{\ensuremath{#1}}
\vspace{0.1cm}}

\newtheorem{definition}{Definition}[section]
\newtheorem{lemma}[definition]{Lemma}
\newtheorem{theorem}[definition]{Theorem}
\newtheorem{corollary}[definition]{Corollary}
\newtheorem{proposition}[definition]{Proposition}

\newtheorem{example}[definition]{Example}

\bgroup

 \newenvironment{myenumerate}
              {\begin{enumerate}\vspace{-1pt}\topsep0pt\parskip0pt\partopsep0pt\itemsep0pt\leftmargin0pt\itemsep2pt\labelwidth0pt\labelsep3pt}
               {\vspace{-1pt}\end{enumerate}} 
  
\newenvironment{mytable}
               {\begin{table}}{\end{table}}
 \newenvironment{mytableT}
               {\begin{table}[t]\vspace{-3pt}}{\vspace{-3pt}\end{table}}

  \newenvironment{myexample}
               {\begin{example}\vspace{3pt}}{\end{example}\vspace{3pt}}              
\newcommand{\acapo}{\\ \indent}
\newcommand{\mysection}[1]{\vspace{-0.6mm}\section{#1}\vspace{-0.6mm}}
                
                            \newenvironment{myproposition}[1]
               {\begin{proposition}[#1]\vspace{1pt}}{\end{proposition}\vspace{1pt}}

\newif\ifmc
\mcfalse 
\mctrue 

\newcommand{\labelx}[1]{\label{#1}}

\newcommand{\vs}{\zeta}
\newcommand{\el}[1]{\ensuremath{|#1|}}
\newcommand{\Mp}{M}
\newcommand{\Np}{N}

\newcommand{\dle}[3]{\ensuremath{#1^{#2}}}

\newcommand{\Us}{\ensuremath{\mathtt{U}}}
\newcommand{\lev}{\ensuremath{\ell}}
\newcommand{\sleq}{\ensuremath{\leq}}

\newcommand{\sls}{\ensuremath{\mathcal{S}}}

\newcommand{\rtsyntax}[1]{{\ensuremath{#1}}}
\newcommand{\ptilde}[1]{{\ensuremath{#1}}}

\newcommand{\kf}[1]{\textup{\textsf{#1}}\xspace}

\newcommand{\uu}{\ensuremath{u}}
\newcommand{\Ia}{\ensuremath{a}}

\newcommand{\Ias}{\ensuremath{\alpha}}
\newcommand{\Ib}{\ensuremath{b}}
\newcommand{\Ibs}{\ensuremath{\beta}}
\newcommand{\y}{\ensuremath{y}}
\newcommand{\PP}{\ensuremath{P}}
\newcommand{\Q}{\ensuremath{Q}}

\newcommand{\DD}{\ensuremath{D}}

\newcommand{\si}[2]{\ensuremath{#1[#2]}}

\newcommand{\swl}{\ensuremath{\s}}

\newcommand{\siil}{\si{\s}{\p}}

\newcommand{\siql}{\si{\s}{\q}}

\newcommand{\cc}{\ensuremath{c}}

\newcommand{\pset}{\ensuremath{\Pi}}

\newcommand{\outP}[3]{\ensuremath{#1!\langle #2,#3\rangle}}
\newcommand{\outDec}[4]{\ensuremath{#1!\langle #2,#3\rangle.#4}}
\newcommand{\outDecl}[4]{\ensuremath{#1!^\lev\langle #2,#3\rangle.#4}}
\newcommand{\outDecp}[3]{\ensuremath{#1!\langle #2,#3\rangle}}

\newcommand{\outs}[4]{\ensuremath{#1!\langle #3,#2\rangle.#4}}
\newcommand{\e}{\ensuremath{e}}
\newcommand{\inpDec}[4]{\ensuremath{#1?( #2,#3).#4}}

\newcommand{\inp}[4]{\ensuremath{#1?( #3,#2).#4}}
\newcommand{\inpl}[4]{\ensuremath{#1?^\lev( #3,#2).#4}}
\newcommand{\inplact}[3]{\ensuremath{#1?( #3,#2)}}

\newcommand{\x}{\ensuremath{x}}
\newcommand{\xl}{\ensuremath{x^\lev}}

\newcommand{\participant}[1]{\ensuremath{\mathtt{#1}}}
\newcommand{\q}{\ensuremath{\participant{q}}}
\newcommand{\p}{\ensuremath{\participant{p}}}

\newcommand{\sdl}[4]{\ensuremath{#1!^\lev\langle\! \langle#3,#2\rangle \!\rangle .#4}}
\newcommand{\sdlpr}[4]{\ensuremath{#1!^{\lev}\langle\! \langle#3,#2\rangle \!\rangle .#4}}

\newcommand{\rdl}[4]{\ensuremath{#1?^{\lev}(\!(#3,#2)\!).#4}}

\newcommand{\z}{\ensuremath{z}}
\newcommand{\pc}{\ensuremath{\ | \ }}
\newcommand{\s}{\ensuremath{s}}
\newcommand{\X}{\ensuremath{X}}

\newcommand{\Xsignature}{\ensuremath{\X(\at{\x}, \alpha)}}
\newcommand{\Xsignaturep}{\ensuremath{\X(\at{x}, \alpha)}}
\newcommand{\Ddef}{\ensuremath{\Xsignature=\PP}}
\newcommand{\Ddeflp}{\ensuremath{\Xsignaturep=\PP}}

\newcommand{\defD}{\ensuremath{\kf{def}\ \DD \ \kf{in}\ }}
\newcommand{\DdefD}{\ensuremath{\kf{def}\ \Ddef \ \kf{in}\ }}
\newcommand{\DdefDlp}{\ensuremath{\kf{def}\ \Ddeflp \ \kf{in}\ }}

\newcommand{\proccall}[3]{\ensuremath{#1\langle\ptilde{#2},\ptilde{#3}\rangle}}
\newcommand{\proccallw}[3]{\ensuremath{#1\langle\ptilde{#2},\ptilde{#3}\rangle}}

\newcommand{\indexed}[4]{\ensuremath{\{#1_#3 : #2_#3\}_{#3 \in #4}}}

\newcommand{\anglep}[2]{\ensuremath{\langle #1, #2\rangle}}
\newcommand{\valheap}[3]{\ensuremath{( #3,\pset,#1 )}}

\newcommand{\delheap}[3]{\ensuremath{(#3,{#2},#1 )}}
\newcommand{\labheap}[3]{\ensuremath{( #3,\pset,#1 )}}

\newcommand{\ifthenelse}[3]{\ensuremath{\kf{if}\ #1\ \kf{then}\ #2\ \kf{else}\ #3}}

\newcommand{\inact}{\ensuremath{\mathbf{0}}}
\newcommand{\nuc}[2]{\ensuremath{(\nu #1)#2}}
\newcommand{\AND}[2]{\ensuremath{#1\ \kf{and}\ #2}}
\newcommand{\NOT}[1]{\ensuremath{\kf{not}\ #1}}

\newcommand{\true}{\kf{true}}
\newcommand{\false}{\kf{false}}
\newcommand{\h}{\ensuremath{h}}
\newcommand{\hH}{\ensuremath{H}}
\newcommand{\K}{\ensuremath{K}}
\newcommand{\mg}{\ensuremath{m}}
\newcommand{\va}{\ensuremath{v^\lev}}

\newcommand{\vap}{\ensuremath{v^{\lev}}}
\newcommand{\at}[1]{\ensuremath{\ptilde{#1}}}
\newcommand{\atw}[1]{\ensuremath{\ptilde{#1}}}

\newcommand{\Par}{\ensuremath{ , }}
\newcommand{\cas}{\ensuremath{r}}
\newcommand{\eq}{\ensuremath{\epsilon}}

\newcommand{\rederr}[3]{\ensuremath{\begin{array}{ll}\kf{if}\ {#1}&\kf{then}\ {#2\Mredsym#3}\\ &\kf{else}\ {#2\error}\end{array}}}
\newcommand{\redsym}{\ensuremath{\longrightarrow}}
\newcommand{\Mredsym}{\ensuremath{\multimap\!\rightarrow}}  

\newcommand{\Mredsymstar}{\ensuremath{{\multimap\!\!\rightarrow^*}}}

\newcommand{\red}[2]{\ensuremath{#1\redsym #2}}
\newcommand{\Mred}[2]{\ensuremath{#1\Mredsym #2}}
\newcommand{\Mredbin}[2]{\ensuremath{#1\Mredsym\, #2}} 
\newcommand{\Mterm}[2]{\ensuremath{#2^{\rceil#1}}}
\newcommand{\mlev}{\ensuremath{\mu}}
\newcommand{\redm}[2]{\ensuremath{#1\redsym^* #2}}

\newcommand{\set}[1]{\ensuremath{\{#1\}}}
\newcommand{\sub}[2]{\ensuremath{\{#1/#2\}}}

\newcommand{\separ}{\ensuremath{~\boldsymbol{|\!\!|}~ }}

\newcommand{\Implies}{\ensuremath{\quad \Rightarrow \quad }}

\newcommand{\mqueue}[2]{\ensuremath{#1 : #2}}
\newcommand{\emptyqueue}[1]{\mqueue{\s}{\epsilon}}
\newcommand{\queue}{\ensuremath{\h}}

\newcommand{\qcomp}[2]{\ensuremath{#1 \cdot #2}}
\newcommand{\qtail}[1]{\ensuremath{\qcomp{\queue}{#1}}}

\newcommand{\qhead}[1]{\ensuremath{\qcomp{#1}{\queue}}}

\newcommand{\subst}[2]{\ensuremath{\{#1 / #2\}}}

\newcommand{\G}{\ensuremath{G}}

\newcommand{\pro}[2]{\ensuremath{#1\upharpoonright#2}}

\newcommand{\T}{\ensuremath{T}}

\newcommand{\TG}{\ensuremath{{\mathcal{T}}}}

\newcommand{\seltype}{\ensuremath{\oplus^\lev \langle \pset,\{\cu_i:\T_i\}_{i\in
I} \rangle }}
\newcommand{\seltypeaj}{\ensuremath{\oplus^\lev \langle \pset,\{\cu_j:\T'_j\}_{j\in
J} \rangle }}

\newcommand{\seltypeaij}{\ensuremath{\oplus^\lev \langle \pset,\{\cu_i:\T_i\}_{i\in
I} \cup\{l_j:\T'_j\}_{j\in
J} \rangle }}
\newcommand{\seltypepj}{\ensuremath{\oplus^\lev \langle \pset,\{\cu_j:\T_j\} \rangle }}
\newcommand{\seltypepjS}[1]{\ensuremath{\oplus^#1 \langle \pset,\{\cu_j:\T_j\} \rangle }}
\newcommand{\seltypeG}{\ensuremath{\oplus^\lev(\pset,\{\cu_i:\pro{\G_i}\q\}_{i\in I})}}
\newcommand{\seltypeT}{\ensuremath{\oplus^\lev\{\cu_i:\pro{\T_i}\q\}_{i\in I}}}
\newcommand{\seltypeTp}{\ensuremath{\oplus^\lev\{\cu_i:\T_i\}_{i\in I}}}

\newcommand{\branchtype}{\ensuremath{\&^\lev(\p,\{\cu_i:\T_i\}_{i\in I})}}
\newcommand{\branchtypeq}{\ensuremath{\&^\lev(\q,\{\cu_i:\T_i\}_{i\in I})}}
\newcommand{\branchtypeS}[1]{\ensuremath{\&^{#1}(\p,\{\cu_i:\T_i\}_{i\in I})}}
\newcommand{\branchtypeqS}[1]{\ensuremath{\&^{#1}(\q,\{\cu_i:\T_i\}_{i\in I})}}
\newcommand{\branchtypeG}{\ensuremath{\&^\lev(\p,\{\cu_i:\pro{\G_i}\q\}_{i\in I})}}
\newcommand{\branchtypeT}{\ensuremath{\&^\lev\{\cu_i:\pro{\T_i}\q\}_{i\in I}}}
\newcommand{\branchtypeTp}{\ensuremath{\&^\lev\{\cu_i:\T'_i\}_{i\in I}}}
\newcommand{\branchtypes}{\ensuremath{\&^\lev\{\cu_i:\T_i\}_{i\in I}}}
\newcommand{\branchtypesG}{\ensuremath{\&^\lev\{\cu_i:\TG_i\}_{i\in I}}}

\newcommand{\cu}{\ensuremath{\lambda}}

\newcommand{\n}{n}

\newcommand{\cadd}{\ensuremath{\cdot}}

\newcommand{\tsn}[3]{\ensuremath{\lfloor#1~\ddagger~ #2\rfloor(#3)}}
\newcommand{\lbrancht}[2]{\ensuremath{#1 \&
(#2, \set{\cu_i:\tsn\T\y{\PP_i}}_{i\in I})}}

\newcommand{\seltypess}{\ensuremath{\oplus^\lev\langle\pset,\{\cu_i:\T_i\}_{i\in I}\rangle}}

\renewcommand{\r}{\ensuremath{r}}

\newcommand{\Sp}{\ensuremath{\mathtt{S}}}

\newcommand{\error}{\ensuremath{\Large{\mathbf\dagger}}}

\newcommand{\nsi}[2]{\ensuremath{\bar{#1}[#2]}}  

\newcommand{\saS}[4]{\ensuremath{#1[#2](#3).#4}}
\newcommand{\sa}[4]{\saS{#1}{#2}{#3}{#4}}   
\newcommand{\lselPS}[4]{\ensuremath{#1 \oplus^{\lev} \anglep{#2}{#3}.#4}}

\newcommand{\lselP}[4]{\lselPS{#1}{#2}{#3}{#4}}
\newcommand{\lselPp}[4]{\lselPS{#1}{#2}{#3}{#4}}
\newcommand{\lsel}[3]{\lselP{#1}{\pset}{#2}{#3}}
\newcommand{\lselp}[3]{\lselPp{#1}{\pset}{#2}{#3}}

\newcommand{\levselp}[4]{\ensuremath{#1 \oplus^{#4} \anglep{#2}{#3}}}

\newcommand{\lbranchPS}[4]{\ensuremath{#1 \&^\lev ({#2},#3)}}

\newcommand{\lbranchP}[3]{\lbranchPS{#1}{#2}{#3}{\lev}}
\newcommand{\lbranchPp}[3]{\lbranchPS{#1}{#2}{#3}{\lev}}
\newcommand{\lambdabranch}[2]{\lbranchP{#1}{#2}{\indexed{\cu}{\PP}{i}{I}}}

\newcommand{\lambdabranchp}[2]{\lbranchPp{#1}{#2}{\indexed{\cu}{\PP}{i}{I}}}
\newcommand{\branch}[3]{\ensuremath{#1 \&^\lev ({#2},#3)}}

\newcommand{\sS}{\ensuremath{s}} 
\newcommand{\sSp}{\ensuremath{s}}
\newcommand{\sSe}[1]{\ensuremath{s}}

\newcommand{\stdqueuep}{\mqueue{\sSp}{\queue}}

\newcommand{\qappendp}[1]{\mqueue{\sSp}{\qtail{#1}}}

\newcommand{\qpopp}[1]{\mqueue{\sSp}{\qhead{#1}}}

\newcommand{\conf}{\ensuremath{C}}
\newcommand{\config}[2]{\ensuremath{< #1 \;\Par\; #2 >}}
\newcommand{\parcon}{\ensuremath{\Vert}}
\newcommand{\proL}[2]{\ensuremath{#1 \Downarrow #2}}
\newcommand{\procalL}[1]{\ensuremath{#1 \Downarrow {\calL}}}

\newenvironment{frcseries}{\fontfamily{frc}\selectfont}{frcursive}

\newcommand{\RLG}{\ensuremath{\calR^\calL_\circ}}
\newcommand{\RL}{\ensuremath{\calR^\calL}}

\newcommand{\proc}{\mbox{${\cal P}\kern-1.5pt\textit{r}$}}

\newcommand{\loweq}{ =_{\calL} }

\newcommand{\lowequiv}{ \simeq_{\calL} }

\newcommand{\meet}{\sqcap}
\newcommand{\join}{\sqcup}

\newcommand{\calL}{\ensuremath{\mathcal L}}

\newcommand{\calR}{\ensuremath{\mathcal R}}

\newcommand{\calS}{\ensuremath{\mathcal S}}

\newcommand{\valheapq}[2]{\ensuremath{( #2,\q,#1 )}}

\newcommand{\Qu}{\ensuremath{{\mathbf{Q}}}}

\newcommand{\lsecure}{\calL--secure}   
\newcommand{\lsecurity}{\calL--security}

\providecommand{\event}{EXPRESS 2011} 

\title{Information Flow Safety in Multiparty Sessions}
\author{Sara Capecchi
\institute{Dipartimento di Informatica\\ Universit\`a di Torino,
corso Svizzera 185, 10149 Torino, Italy\thanks{Work partially funded by the ANR-08-EMER-010 grant PARTOUT, and by the MIUR Projects DISCO and IPODS.}}
\email{capecchi@di.unito.it}
\and
 Ilaria Castellani 
\institute{INRIA\\ 2004 route des Lucioles, 06902 Sophia Antipolis, France}
\email{\quad ilaria.castellani@inria.fr} 
\and Mariangiola Dezani-Ciancaglini
\institute{Dipartimento di Informatica\\ Universit\`a di Torino,
corso Svizzera 185, 10149 Torino, Italy}
\email{dezani@di.unito.it}
}

\def\titlerunning{Information flow safety in multiparty sessions}
\def\authorrunning{S. Capecchi, I Castellani \& M. Dezani-Ciancaglini}
\begin{document}

\maketitle

\pagestyle{plain}

\begin{abstract} 
  We consider a calculus for multiparty sessions enriched with
  security levels for messages.  We propose a monitored semantics for
  this calculus, which blocks the execution of processes as soon as
  they attempt to leak information.  
We illustrate the use of our monitored
  semantics with various examples, and show that the induced
safety property
implies a noninterference
  property studied previously.
\end{abstract}


\mysection{Introduction}
\label{introd}
With the advent of web technologies, we are faced today with a powerful
computing environment which is inherently parallel, distributed and
heavily relies on 
communication.  
Since computations take place concurrently on several heterogeneous
devices, controlled by parties which possibly do not trust each other,
security properties such as confidentiality and integrity of data
become of crucial importance.
\acapo A {\em session} is an abstraction for various forms of
``structured communication'' that may occur in a parallel and
distributed computing environment. Examples of sessions are a
client-service negotiation, a financial transaction, or an interaction
among different services within a web application.  {\em Session
types}, which specify the expected behaviour of participants in
sessions, were originally introduced in \cite{THK}, on a variant of
the $\pi$-calculus \cite{MilnerR:commspc} including a construct for
session creation and two $n$-ary operators of labelled internal and
external choice, called selection and branching. The basic properties
ensured by session types are the absence of communication errors
(communication safety) and
the conformance to the session protocol (session fidelity).
Since then, more powerful {\em session calculi} have been investigated, allowing
delegation of tasks among participants and multiparty interaction
within a single session, and equipped with
increasingly sophisticated session types, ensuring additional
properties like progress.
\acapo In previous work \cite{CCDR10}, we addressed the question of 
incorporating security requirements into session types.
To this end, we considered a calculus for multiparty
sessions with delegation, enriched with security levels
for both session participants and data.
We proposed a session type system for this calculus,
adding 
access control and secure information flow requirements 
in the typing rules in order to guarantee the preservation of data {\em
confidentiality} during session execution.
\acapo In this paper, we move one step further by equipping the above calculus with a {\em
monitored semantics}, 
which blocks the execution of processes as soon as
they attempt to leak information, raising an error. Typically, this happens when a
process tries to participate in a public 
communication after receiving or
testing a secret value.
This monitored semantics induces a natural notion of safety on
processes: a process is safe if 
all its monitored computations are successful
(in a dynamically evolving environment and in the presence of a
passive attacker, which may only change secret information at each step).
 \acapo Expectedly, this monitored semantics is closely
related to the security type system presented
in~\cite{CCDR10}. Indeed, some of the constraints imposed by the
monitored operational rules are simply lifted from the 
typing rules.
However, there are two respects in which the constraints of the
monitoring semantics are simpler.
First, they refer to individual computations. In other words, they
are {\em local} whereas type constraints are both local and {\em global}.
Second, one of these constraints (the lower bound for the level
of communications in a
given session) may be dynamically computed during execution, and
hence services do not need to be statically annotated with levels,
as was required by the type system of~\cite{CCDR10}. This means that the language itself
may be simplified when the concern is on safety rather than typability.
\acapo Other advantages of safety over typability are not
 specific to session calculi. 
Like security, safety is a semantic notion. Hence it is {\em
 more permissive} than typability in that it ignores unreachable parts
of processes: for instance, in our setting, a
 high conditional with a low branch will be ruled out by the type
 system but will be considered safe if the low branch is never taken.
Safety also offers
more {\em flexibility} than types in the context of web
programming, where security policies may change dynamically.
%
\acapo 
Compared to security, safety has again the advantage of locality
versus globality.
In session calculi, it also improves on security in another respect.
Indeed, in these calculi processes communicate asynchronously
and messages transit in queues before being consumed by their
receivers. Then, while the monitored semantics blocks the very act of
putting a public message in the queue after a secret message has been
received, 
a violation of the security property can
only be detected {\em after} the public message has been put in the
queue, that is, after the confidentiality breach has occurred,
and possibly already caused damage. 
This means that safety allows
{\em early} leak detection, whereas security only allows
{\em late} detection.
\acapo
Finally, safety seems more appealing than security
when the dangerous behaviour comes from
an accidental rather than an intentional transgression of the security
policy.  Indeed, in this case a monitored semantics could offer useful
feedback to the programmer, in the form of informative error messages.
Although this possibility is not explored in the present paper,
it is the object of ongoing work.
\acapo The main contribution of this work is 
a monitored semantics for a multiparty session calculus,
and the proof that 
the induced 
{\em information flow safety} property strictly implies the {\em information flow
security} property of~\cite{CCDR10}. 
While the issue of safety 
has recently received much attention
in the security community (see Section ~\ref{conclusion}), it has not, to our knowledge,
been addressed in the context of session calculi so far.
\acapo The rest of the paper is organised as follows. In
Section~\ref{introex} we motivate our approach with an
example. Section~\ref{synsem} introduces the syntax and semantics of
our calculus.  In Section~\ref{MonSemSec} we recall the definition 
of security from~\cite{CCDR10} and illustrate it with examples.
Section~\ref{monsem} presents our 
monitored semantics and Section~\ref{MonSemSaf} introduces our notion of
safety and establishes its relation to security.
Finally, Section~\ref{conclusion} concludes with a discussion
on related and future work. 
\mysection{Motivating example}
\label{introex}
Let us illustrate our approach with an introductory example, inspired
by~\cite{Mitchell}.  Suppose we want to model the interaction between
an online health service $\Sp$ and a user $\Us$.  Each time the user
wishes to consult the service, she opens a connection with the server
and sends him her username (here by convention we shall use ``she''
for the user and ``he'' for the server), assuming she has previously
registered with the service.  She may then choose between two kinds of
service:
\begin{enumerate}
	\item \textsl{simple consultation}: the user asks questions
          from the medical staff. Questions and answers are public, for
          instance they can be published in a forum visible to every
          user. The staff has no privacy constraint.
	\item \textsl{medical consultation}: the user sends questions
          together with private data (e.g., results of medical exams)
          to the medical staff, in order to receive a diagnosis or
          advice about medicines or further exams. To access these
          features she must enter a password, and wait for a secure
          form on which to send her data.  Here questions and answers
          are secret (modelling the fact that the they are sent in a
          secure mode and that the staff is compelled to maintain
          privacy).
\end{enumerate}
More precisely, this interaction may be described by the following
protocol, in which we add the possibility that the user accidentally
reveals her private information:
\begin{enumerate}
\item  \Us\ opens a connection with \Sp\ and sends her  username to \Sp;
\item \Us\ chooses between Service 1 and Service 2;
\item[3.a] Service 1:  \Us\ sends a question to \Sp\ 
and waits for an answer; 
\item[3.b] Service 2: \Us\ sends her password to \Sp\ and waits for a
  secure form. She then sends her question and data on the form and
  waits for an answer from \Sp.  A reliable user will 
use the form correctly and send data in a secure mode.
Instead, an unreliable user will forget to use the
  form,  or use it wrongly, thus leaking some of her private data.
 This may result in private information being sent to a public forum or to medical
 staff which is not compelled to maintain privacy.
\end{enumerate}
In our calculus, 
this scenario may be
described as the parallel composition of the processes in Figure~\ref{medical-example}.
\begin{figure}
\begin{tabular}{l}
\begin{tabular}{lcll}
  \participant{I} &  $=$ &\nsi{a}{2} &
  \\[2mm]
 \Us & $ = $& $a[1](\alpha_1).\outDec{\alpha_1}{2}{\mathsf{un}^{\bot}}{}$ &  $\kf{if} \ \mathit{\textsl{simple}^\bot}$ 
\ \kf{then}\ \levselp{\alpha_1}{2}{\mathbf{sv1}}{\bot}.\outDec{\alpha_1}{2}{\mathsf{que}^{\bot}}{\inp{\alpha_1}{{\mathsf{\mathit{ans}}}^\bot}{1}{}} \inact\\
&&&  \hspace{52pt}$  \kf{else} \
\levselp{\alpha_1}{2}{\mathbf{sv2}}{\bot}. \outDec{\alpha_1}{2}{\mathsf{pwd}^{\top}}{\inp{\alpha_1}{{\mathit{form^{\top}}}}{2}{}$
\\
 &&&  \hspace{52pt} $  \kf{if} \ \mathit{\textsl{gooduse}(\mathit{form}^{\top})} \ \kf{then} \ \outDec{\alpha_1}{2}{\mathsf{que}^\top}{\inp{\alpha_1}{\mathit{ans}^\top}{2}{\inact}}}$ \\
&&& \hspace{137pt} $\kf{else} \
\outDec{\alpha_1}{2}{\mathsf{{que}}^\bot}{\inp{\alpha_1}{\mathit{ans}^\bot}{2}{\inact}}$
\end{tabular}\\
\\
\begin{tabular}{lcll}
 \Sp & $ = $& $a[2]  (\alpha_2).$ & $\inp{\alpha_2}{\mathit{un}^{\bot}}{1}{}$  \\
 &&&\ $ \alpha_2 \ \&^\bot (1, \{ \mathbf{sv1}: \inp{\alpha_2}{\mathit{que}^{\bot}}{1}{\outDec{\alpha_2}{1}{\mathsf{ans}^\bot}{\inact}} ,$\\
& & & $\qquad \qquad \quad\!\! \mathbf{sv2}:
\inp{\alpha_2}{\mathit{pwd}^\top}{1}{\outDec{\alpha_2}{1}{{\mathsf{form}^\top}}
{\inp{\alpha_2}{\mathit{que}^\top}{1}{\outDec{\alpha_2}{1}{\mathsf{ans}^\top}{\inact}}}
}\}$\\
\end{tabular}
\end{tabular}
\smallskip
\caption{The online medical service example.}\label{medical-example}
\smallskip
\end{figure}
%
%
%
\acapo A session is an activation of a service, involving a number of
participants with predefined roles.  Here processes \Us\ and \Sp\
communicate by opening a session on service $a$. The initiator
$\nsi{a}{2}$ specifies that the number of participants is 2.  
Participants are denoted by integers: here \Us=1, \Sp=2.
In process \Us, the prefix $\Ia[1](\alpha_1)$ means that \Us\
wants to act as participant 1 in service \Ia, using channel $\alpha_1$ to
communicate. Dually, in \Sp, $\,\Ia[2](\alpha_2)$ means that  \Sp\ will
  act as participant 2 in service \Ia, communicating via channel $\alpha_2$.
\\
\indent 
Security levels appear as superscripts on both data and some operators (here $\bot$
means ``public'' and $\top$ means ``secret''): the user name
\textsf{un} 
and the message
contents in Service 1 can be public; the password \textsf{pwd} and the
information exchanged in Service 2 should be secret.
Levels on the operators are needed to track indirect flows,
as will be explained in Section~\ref{MonSemSaf}. They may be ignored for the time being.
\acapo When the session is established, via a synchronisation between the
initiator and the prefixes $\Ia[i](\alpha_i)$, \Us\ sends to \Sp\ her
username $\mathsf{un}^{\bot}$. Then, according to whether she wishes a
simple consultation or not, she chooses between the two 
services $\mathbf{sv1^\bot}$ and $\mathbf{sv2^\bot}$. 
This choice is expressed by the internal choice construct $\oplus$: $\kf{if} \
\mathit{\textsl{simple}^\bot} \ \kf{then} \
\levselp{\alpha_1}{2}{\mathbf{sv1}}{\bot}\ldots\ \kf{else} \
\levselp{\alpha_1}{2}{\mathbf{sv2}}{\bot} \ldots\ $ describes a
process sending on $\alpha_1$ to participant 2  either  label
\textbf{sv1} or \textbf{sv2}, depending on the  value of $\textsl{simple}^\bot$. If \Us\
chooses $\mathbf{sv1}$,
then she sends to \Sp\ a question (construct
\outDecp{\alpha_1}{2}{\mathsf{que}^{\bot}}), receives the answer
(construct \inplact{\alpha_1}{{\mathsf{\mathit{ans}}}^\bot}{1}{}).
If \Us\ chooses $\mathbf{sv2}$, then she sends her
password to \Sp\ and then waits to get back a secure form. At this point,
according to her reliability ($\kf{if} \ \mathit{\textsl{gooduse}\,(\textit{form}^\top)}$) she
either sends her question and data in the secure
form, or wrongly sends them in clear. 
The difference
between the secure and insecure exchanges is modelled by the different
security levels tagging 
values and variables in the prefixes
$\outP{\alpha_1}{2}{\mathsf{que}^{\top / \bot}}$ and
$\inplact{\alpha_1}{\mathit{ans}^{\top /\bot }}{2}$.
\acapo Dually, process \Sp\ receives the username from \Us\ and then waits
for her choice of either label $\mathbf{sv1}$ or label
$\mathbf{sv2}$.  This is described by the external choice operator $\&$: $\&^\bot (1, \{ \mathbf{sv1}:  \ldots ,
 \mathbf{sv2}:\ldots )$ expresses the reception of the label
 \textbf{sv1} or of the label \textbf{sv2} from participant 1.  In
the first case, \Sp\ will then receive a question and
send the answer. This whole interaction is public. In the second case, \Sp\
receives a password and sends a form, and then receives a question and sends the answer. In
this case the interaction is secret.
\acapo Note that the execution of process $\participant{I}\pc \Sp \pc \Us$
may be insecure if \Us\ is unreliable.  Indeed, in \Us's code, the
test on $\mathit{\textsl{gooduse}\,(\textit{form}^\top)}$ 
uses the secret value $\mathit{form}^\top$.
Now, for security to be
granted in the subsequent execution, all communications depending on
$\,\mathit{form}^\top$ should be secret.  However, this will not be the
case if the second branch of the conditional is taken, since in this
case \Us\ sends a public question.
On the other hand, the execution is
secure when the first service is used, or when the second service is
used properly.
\acapo
This process is rejected by the type system of~\cite{CCDR10},
  which must statically ensure the correction of all
  possible executions.  Similarly, the security property
  of~\cite{CCDR10} fails to hold for this process, since
two different public behaviours may be exhibited 
after testing the secret value $\mathit{\textsl{gooduse}\,(\textit{form}^\top)}$:
in one case the empty behaviour, in the other
  case the emission of $\mathsf{que}^\bot$.  Moreover, the
  bisimulation used to check security will fail only once
  $\mathsf{que}^\bot$ has been put in the queue, and thus
possibly exploited by an attacker. By contrast, the monitored
  semantics will block the very act of putting $\mathsf{que}^\bot$ in
  the queue.
   \acapo 
  For the sake of conciseness, we deliberately simplified the scenario in the
  above example, by using finite services and a binary session between a server and a
  user. 
Note that several such sessions could run in parallel, each
  corresponding to a different impersonation of the user. A more
  realistic example would involve persistent services and allow several users to interact within the
  same session, and the server to delegate the question handling to
  the medical staff.
  This would bring into the scene other important features of our
  calculus, namely multiparty interaction and the mechanism of
  delegation. Our simple example is mainly meant to highlight the
  novel issue of monitored execution.
\mysection{Syntax and Standard Semantics}
\label{synsem}
Our calculus is essentially the same as that studied
in~\cite{CCDR10}. 
For the sake of simplicity,
we do not consider here access control and declassification, 
although their addition would not pose any problem. 
\acapo
Let $(\sls,\sleq)$ be a finite lattice of {\em security levels},
ranged over by $\ell,\ell'$. We denote by $\join$ and $\meet$ the join
and meet operations on the lattice, and by $\bot$ and $\top$ its
minimal and maximal elements.
We assume the following sets: 
{\em values} (booleans, integers), ranged over by $v, v'\ldots$,
{\em value variables}, ranged over by $\x,\y \ldots$,
{\em service names}, ranged over by $\Ia,\Ib,\dots$, each of which has an {\em arity}
$n\geq 2$ (its number of participants), {\em service name variables},
ranged over by $\vs,\vs',\dots$,
{\em identifiers}, i.e.,
  service names and value variables, ranged over by $\uu,w,\dots$, 
\emph{channel variables},
  ranged over by $\alpha,\beta,\dots$, 
 and \emph{labels}, ranged
  over by $\lambda,\lambda',\dots$ (acting like 
  labels in labelled records).  
{\em Sessions}, the central abstraction of our calculus, are denoted
with $s, s'\ldots$. A session represents a particular instance or
activation of a service. Hence sessions only appear at runtime.  We
use $\participant{p}$, $\participant{q}$,\ldots to denote the {\em
  participants} of a session. In an $n$-ary session (a session
  corresponding to an $n$-ary service) $\participant{p}$,
  $\participant{q}$ are assumed to range over the natural numbers
  $1,\dots,n$. We denote by \pset\ a non empty set of participants.
Each session $s$ has an associated set of \emph{channels with role}
\si{\s}{\p}, one for each participant.  Channel \si{\s}{\p} is the
private channel through which participant \p{} communicates with the
other participants in the session \s.  A new session $s$ on an $n$-ary
service $\Ia$ is opened when the {\em initiator} $\nsi{\Ia}{\n}$
of the service synchronises with $n$ processes of the form
$\sa{\Ia}{1}{\Ias_1}{\PP_1}, \ldots, \sa{\Ia}{n}{\Ias_n}{\PP_n}$, 
whose channels $\alpha_\p$ then get replaced by $\si{\s}{\p}$
in the body of $P_\p$.
{While binary sessions may often be viewed as an interaction between a
user and a server, multiparty sessions do not exhibit the same
asymmetry. This is why we use of an initiator to start the session once all
the required ``peer'' participants are present.}
We use \cc\ to range over channel
variables and channels with roles.  Finally, we assume a set of
\emph{process variables} $\X,Y,\dots$, in order to define recursive
behaviours.
\label{sec:syntax}
\begin{table}
\footnotesize{
\centering
\begin{tabular}{lll}
\begin{tabular}{rclr}
\r & ::= &  $\Ia$\ \separ $\rtsyntax{\s}$ & Service/Session Name \\[1mm]
\cc & ::= & \Ias \ \separ \rtsyntax{\si\s\p}& {Channel}
\\[1mm]
\uu & ::= & $\vs$ \separ $\Ia$ &{Identifier}\\[1mm]
$v$   & ::= & $\true$ \separ $\false$ \separ \ \ldots&{Value}\\[1mm]
\e   & ::= & $\xl \separ  \va  \, \separ \NOT{\e} $\\
&\separ&\AND{\e}{\e'}\separ \ldots
&{Expression}
\\[1mm]
\DD   & ::= & \Ddef &{Declaration}
\\[2mm]
\pset & ::= & \set\p\ \separ $\pset\cup\set\p$ &{Set of participants}\\\\
$\vartheta$ & ::= & \rtsyntax{v^\lev \ \separ \si\s\p^\lev \ \separ \lambda^\lev} & {Message content}
\\[2mm]
\mg   & ::= & \rtsyntax{\valheap{\mbox{$\vartheta$}}{\pset}{\p}}  \quad \quad \quad &{Message in transit}
\\[1mm]
\h   & ::= & \rtsyntax{{\at{\mg}\,\cdot \h}  \ \separ
  {\eq}}&Queue
\\[1mm]
  \hH   & ::= & \rtsyntax{\hH\cup\set{ \sS :\h} \ \separ \rtsyntax{\emptyset}}
  &\Qu-set
\\[1mm]
   \end{tabular}
&
&
\begin{tabular}{rclr}
 \PP & ::=  & \nsi{\uu}{\n}   &   {$n$-ary session initiator}\\
     & \separ & \sa\uu\p\Ias\PP   &   {\p-th session participant}\\
& \separ & \outDec{\cc}{\pset}{\e}\PP & {Value send}\\
& \separ & \inpDec{\cc}{\p}{x^{\lev}}\PP & {Value receive}\\
& \separ & \outDecl{\cc}{\pset}{\uu}\PP & {Service name send}\\
& \separ & \inpl{\cc}{\vs}{\p}\PP & { Service name receive}\\
      & \separ & \sdl{\cc}{\cc'}{\q}{\PP} & {Channel send}\\
      & \separ & \rdl{\cc}{\Ias}{\p}{\PP} &{Channel receive}\\
      & \separ & \lsel{\cc}{\lambda}{\PP} &{Selection}\\  
      & \separ & \lambdabranch{\cc}{\p} & {Branching}\\
            & \separ & \ifthenelse{\e}{\PP}{\Q} &{Conditional}\\
      & \separ & \PP \pc \Q  &{Parallel}\\
      & \separ & \inact & {Inaction}\\
     & \separ & \nuc{\Ia}{\PP} & {Name hiding}\\
             & \separ & \defD\PP & {Recursion}\\
      & \separ & \proccall{\X}{\e}{\cc} & {Process call}
\\
\end{tabular}
 \end{tabular}}
\caption{Syntax of processes, expressions and  queues.}\label{tab:syntax}
\end{table}
\acapo
As in \cite{CHY07}, in order to model TCP-like asynchronous
communications (with non-blocking send but message order preservation
between a given pair of participants), we use {\em queues of
  messages}, denoted by \h; 
an element of $h$ may be one of the following: a value message 
\valheap{v^\lev}{\pset}{\p}, indicating that the value
$v^\lev$ is sent by participant \p{} to all
participants in $\pset$; a service name message 
$\valheap{a^\lev}{\pset}{\p}$, with a similar meaning;
a channel
message \delheap{\si\s{\p'}^\lev}{\q}{\p}, indicating that $\p$
delegates to $\q$ the role of $\p'$ with level $\ell$ in session
$\s$; and a label message \labheap{\mbox{$\lambda^\lev$}}{\q}{\p},
indicating that $\p$ selects the process with label $\lambda$ among
those offered by the set of participants $\pset$.  The empty
queue is denoted by \eq, and the concatenation of a message $\mg$
to a queue $\h$ by $\h \cadd \mg$.  Conversely, $\mg \cadd \h$ means
that $\mg$ is the head of the queue.  Since there may be interleaved, nested and
parallel sessions, we distinguish their queues with names.  We
denote by $\swl: \h$ the {\em named queue} \h{} associated with
session \s.  We use $H,K$ to range over sets of named queues with {different session names,} also
called \Qu-sets. 
\acapo
Table ~\ref{tab:syntax} summarises the syntax of \emph{expressions}, ranged over by $\e,\e',\dots$, and
of \emph{processes}, ranged over by $\PP,\Q\dots$, 
as well as the {\em runtime syntax} of the calculus (sessions,
channels with role, messages, queues).
\acapo Let us briefly comment on the primitives of the
  language. We already described session initiation.
Communications within a session
are performed on a channel 
using the next four pairs of primitives: 
the send 
and receive 
of a value; the send and receive of a
service name; the send and receive of a channel (where one participant
transmits to another the capability of participating in another
session with a given role) and
the selection 
and branching  
operators (where one participant chooses
one of the branches offered by another participant). 
Apart from the value send and receive constructs,
all the send/receive and choice primitives are decorated with
  security levels, whose use will be justified later.
When there is no risk of confusion we will omit the set delimiters $\{,\}$,
particularly around singletons. \acapo
\begin{mytableT}
{
\centering
\begin{tabular}{c}
 $\sa a {1}{\alpha_1}{\PP_1}
        \pc...\pc\sa a {n}{\alpha_n}{\PP_n} \pc \nsi{a}{n}$
         $\redsym (\nu {\swl})\config{
        \PP_{1}\sub{\si\s {1}}{\alpha_{1}} \pc ...\pc
        \PP_n\sub{\si \s {n}}{\alpha_n}}{ \sS : \eq}$  \\
        \hfill [Link]
\\[2mm]
\red{\config{\outDec{\siil}{\pset}{\e}{\PP}}{ \mqueue{\sSp}{\queue}}}
    {\config{\PP }{\mqueue{\sSp}{\qtail{\valheap{\dle{v}{\lev}{\lev'} }{\p}{\p}} } } }
       where \at{\e}$\downarrow$\at{\va} \\
        \hfill  [SendV] 
\\[2mm]
$\config{\inpDec{\siql}{\p}{\x^{\lev}}{\PP} }{\qpopp{\valheapq{\dle{v}{\lev}{\lev'}}{\p}}}$
    $\redsym  \config{\PP\subst{v}{\x} } {
    \mqueue{\sSp}{\queue}}$\ \ \ \ \ \  \\
        \hfill [RecV]
\\[2mm]
\red{\config{\outDecl{\siil}{\pset}{\Ia}{\PP}}{ \mqueue{\sSp}{\queue}}}
    {\config{\PP }{\mqueue{\sSp}{\qtail{\valheap{\Ia^\lev }{\p}{\p}} } } }
        \\
        \hfill  [SendS] 
\\[2mm]
  $\config{\inpl{\siql}{\vs}{\p}{\PP} }{\qpopp{\valheapq{\Ia^\lev}{\p}}}$
    $\redsym  \config{\PP\subst{\Ia}{\vs} } {
    \mqueue{\sSp}{\queue}}$\ \ \ \ \ \  \\
        \hfill [RecS]
\\[2mm]
  \red{\config{    \sdlpr{\siil}{\s'[\p']}{\q}{\PP}   }{ \mqueue{\sSp}{\queue}}}
    {\config{\PP }{ \mqueue{\sSp}{\qtail{\delheap{\si{\s'}{\p'}^{\lev}}{\q}{\p}}}}}\\
        \hfill  [SendC] 
\\[2mm]
    \red{\config{    \rdl{\siql}{\Ias
           }{\p}{\PP}   }{ \qpopp{\delheap{\si{\s'}{\p'}^{\lev}}{\q}{\p}}}}
    {\config{\PP\subst{\si{\s'}{\p'}}{\Ias} }{ \stdqueuep}}\\
        \hfill [RecC] 
\\[2mm]
    \red{\config{\lselp{\siil}{\lambda}{\PP}}{ \stdqueuep}}
    {\config{\PP }{ \qappendp{\labheap{\lambda^{\lev}}{\p}{\p}}}}\\
        \hfill  [Label]\\
         \red{ \config{ \branch{\siql}{\p}{\indexed{\cu}{\PP}{i}{I}} } {\qpopp{\valheapq{\lambda_{i_0}^{\lev}}{\p}}}}
{ \config{ \PP_{i_0
} }{ \mqueue{\sSp}{\queue}}$ where $(i_0 \in I)} \\
        \hfill  [Branch]\\[2mm]
  \red{\ifthenelse{\e}{\PP}{\Q}}{\PP}  where $\e \downarrow \true^\lev$
    \quad
\red{\ifthenelse{\e}{\PP}{\Q}}{\Q}  where  $\e \downarrow \false^\lev$\\
        \hfill [If-T, If-F]
\\[2mm]
$\DdefD \proccallw{X}{\e}{\si\s\p}$
    $\redsym \DdefD
    \PP\subst{\at{\va}}{\at{\x}}\subst{\atw{\si\s\p}}{\alpha}$\ \ \ \
    where  $\ptilde{\e}\downarrow \at{\va}$ \\
        \hfill [Def]
\\[2mm]
 \red{\config{\PP}{\hH}}{\nuc{\tilde{\swl}}\config{\PP'}{\hH'}} \Implies\\
 \red{\config{\defD(\PP \pc \Q)}{\hH}}{\nuc{\tilde{\swl}}\config{\defD(\PP' \pc \Q)}{\hH'}}\\
        \hfill [Defin]
\\[2mm]
\red{C}{\nuc{\tilde{\swl}}C'}
\Implies\red{\nuc{\tilde{\r}}(\,C\,\parcon\: C'')}{\nuc{\tilde{\r}}{\nuc{\tilde{\swl}}(\,C'\,\parcon\: C'')}}\\
        \hfill  [Scop]
\end{tabular}
\caption{Standard reduction rules.}\label{tab:reduction2}}
\end{mytableT}
%
The operational semantics consists of a reduction relation on
  configurations $\config{\PP}{\hH}$, which are pairs of a process
  \PP\ and a \Qu-set \hH.  Indeed, queues need to be isolated from
  processes in our calculus (unlike in other session calculi, where
  queues are handled by running them in parallel with processes),
  since they will be the observable part of processes in our security
  and safety notions.
\acapo
A {\em configuration} is a pair $C =\, \config{\PP}{\hH}$ of a process
\PP\ and a \Qu-set \hH, possibly restricted with respect to service
and session names, or a parallel composition {$(\,C \parcon\, C')$} 
{of two configurations whose \Qu-sets have disjoint session names.}
In a configuration \nuc{\swl}{\config{\PP}{\hH}}, all occurrences of
\si\s\p\ in \PP\ and \hH\ and of \swl\ in \hH\ are bound.  By abuse of
notation we often write $\PP$ instead of
\config{\PP}{\emptyset}.\acapo As usual, the operational semantics is
defined modulo a structural equivalence $\equiv$.  The structural
rules for processes are standard~\cite{MilnerR:commspc}. Among the
rules for queues, we have one for commuting independent messages and
another one for splitting a message for multiple recipients. {The
  structural equivalence of configurations {allows
  the parallel composition $\parcon$ to be eliminated via} the rule:

\processes{\nuc{\tilde{\cas}}{\config{\PP}{\hH} } \parcon\, \nuc{\tilde{\cas}'}{\config{\Q}{\K} } \:\: \equiv\:\: 
\nuc{\tilde{\cas}\tilde{\cas}'}{\config{\PP\pc \Q}{\hH \cup \K}}}
\noindent
where by hypothesis the session names in the \Qu-sets \hH, \K\ are disjoint, by Barendregt convention $\tilde{\cas}$ and $\tilde{\cas}'$ have empty intersection and there is no capture of free names, and $\nuc{\tilde{\r}}{\,C}$ stands for
$\nuc{\r_1}{\cdots\nuc{\r_k}{\,C}}$, if $\tilde{r}=r_1\cdots r_k$.}
Note that, modulo 
$\equiv$, 
each configuration has the form $\nuc{\tilde{\r}}{\config{\PP}{\hH}
}$. 
\acapo
The transitions for configurations have the form $\conf\redsym
\conf'$. They are derived using the reduction rules in
Table~\ref{tab:reduction2}, where we write $P$ as short for \config P\emptyset. 
\acapo
Rule [Link] describes the initiation of a
new session among $n$ processes, corresponding to an activation of the
service $\Ia$ of arity $n$.  After the connection, the
participants share a private session name \swl\ and the corresponding
queue, initialised to $ \sS :\eq$.  In each
participant $\PP_\p$,  the channel variable $\alpha_\p$ is replaced by the channel with
role $\si{\s}{\participant{p}}$. This is the only synchronous interaction of the calculus.
All the other communications, which take place within an established session,
are performed asynchronously in two steps, via push and pop operations
on the queue associated with the session. 
\acapo
The output rules [SendV], [SendS], [SendC] 
and [Label] push values, service names, channels and labels, respectively, into the
queue 
$\mqueue{s}{h}$.  In rule [SendV], $\at{\e}\downarrow\at{\va}$ denotes
the evaluation of the expression $\e$ to the value \va, where  $\lev$
is the join of the security levels of the variables and values
occurring in $e$.
\acapo The input rules [RecV], [RecS], [RecC] 
and
[Branch] perform the complementary operations.  
Rules [If-T], [If-F], [Def] and [Defin] are standard.
The contextual rule [Scop] is also standard. In this rule, 
Barendregt convention ensures that the names in $\tilde{s}$
are disjoint from those in $\tilde{r}$ and do not appear in $C''$.
As usual, we use $\redm{}{}$ for the reflexive and transitive
closure of $\red{}{}$.
\acapo We assume that communication safety and  session fidelity are assured by a standard session type system \cite{CHY07}
\mysection{Security}
\label{MonSemSec}
As in \cite{CCDR10}, we assume that the observer can see the 
messages in session queues.  
As usual for security, observation is relative to a given downward-closed set of levels
{$\calL\subseteq\calS$}, the intuition being that an observer who
can see messages of level $\ell$ can also see all messages of level
$\ell'$ lower 
than $\ell$. In the following, we shall always use $\calL$ to denote a
downward-closed subset of levels. For any such $\calL$, an
$\calL$-observer will only be able to see messages whose levels belong
to $\calL$, what we may call {\em $\calL$-messages}.  Hence two queues
that agree on $\calL$-messages will be indistinguishable for an
$\calL$-observer. 
{Let now {\em $\overline{\calL}$-messages} be the
  complementary messages, those the $\calL$-observer cannot see.}
{Then, an $\calL$-observer may also be viewed as}
{an attacker who tries to reconstruct the dependency between
$\overline{\calL}$-messages and $\calL$-messages (and hence,
ultimately, to discover the $\overline{\calL}$-messages),
  by injecting himself different 
  $\overline{\calL}$-messages at each step
and observing their effect on $\calL$-messages.}
\acapo
To formalise this intuition, a notion of $\calL$-equality
$\loweq$ on $\Qu$-sets is introduced, representing
indistinguishability of $\Qu$-sets by an $\calL$-observer. Based on
$\loweq$, a notion of $\calL$-bisimulation $\lowequiv$ formalises
indistinguishability of processes by an $\calL$-observer.  Formally, a
queue $s: h$ is $\calL$-observable if it contains some message with
level in $\calL$. Then two $\Qu$-sets are {\em \calL-equal}\/ if their
$\calL$-observable queues have the same names and contain the same
messages with level in \calL.  This equality is based on an
$\calL$-projection operation on $\Qu$-sets, which discards all
messages whose level is \mbox{not in \calL}.
\acapo
Let the function $lev$ 
be given by:
$
lev(v^{\lev})= lev(a^{\lev})=
lev(\si\s{\p}^{\lev})=lev(\lambda^\lev)=\lev$.

\begin{definition}{{\bf ($\calL$-Projection)}}
\label{def:projP}
The projection operation 
$\procalL{}$ is defined inductively on
messages, queues and $\Qu$-sets as follows: 

\processes{\begin{array}{l}
\proL{(\p, \Pi, \vartheta)}{\calL}=\begin{cases}
    (\p, \Pi, \vartheta)  & \text{if }lev(\vartheta)\in\calL, \\
    \epsilon  & \text{otherwise}
\end{cases}\end{array}\qquad
\begin{array}{l}\proL{\epsilon}\calL=\epsilon\\[2pt]
\proL{(\at{\mg}\,\cdot \h)}\calL=\proL{\mg}\calL\,\cdot \proL\h\calL 
\end{array}
}
\processes{\begin{array}{l}
\proL{\emptyset}\calL=\emptyset\qquad
\proL{(\hH\cup\set{ \sS :\h})}\calL=\begin{cases}
\proL{\hH}\calL\cup\set{ \sS :\proL\h\calL} & \text{if }\proL{h}\calL\neq\epsilon, \\
 \proL{\hH}\calL & \text{otherwise}
\end{cases}
\end{array}
}
\noindent
\end{definition}

\begin{definition}{{\bf ($\calL$-Equality of $\Qu$-sets)}} \text{~}\\
Two $\Qu$-sets $H$ and $K$ are $\calL$-{\em
equal}, written $H\loweq K$, if $\proL\hH\calL=\proL\K\calL$.
\end{definition}

 The idea 
is to test processes by running
  them in conjunction with $\calL$-equal queues. However,
we cannot allow arbitrary combinations of processes with queues,
since this would lead us to reject intuitively secure processes as
simple as $\inp{s[2]}{x^\bot}{1}{\inact}$ and $\outs{s[1]}{\true^\bot}{2}{\inact}$.
As argued in~\cite{CCDR10}, we may
get around this problem by imposing two simple conditions, one on
$\Qu$-sets (\emph{monotonicity}) and the other on configurations
(\emph{saturation}).  These conditions are justified by the fact that
they are always satisfied in initial computations generated by typable
processes (in the sense of~\cite{CCDR10}).
\acapo
The first condition requires that in a $\Qu$-set, the security levels of 
messages with the same sender and common  receivers should
never decrease along a sequence.

  \begin{definition}{{\bf (Monotonicity)}}
  A queue is {\em monotone} if $lev(\vartheta_1)\leq
  lev(\vartheta_2)$ 
whenever the message $(\p, \Pi_1,
  \vartheta_1)$ precedes the message $(\p, \Pi_2, \vartheta_2)$ in the
  queue and $\Pi_1\cap\Pi_2\not =\emptyset$.
 \end{definition}

The second condition requires that in a configuration,
the $\Qu$-set should always contain
enough queues to enable all outputs of the process to reduce.

\begin{definition}{{\bf (Saturation)}} 
 A configuration $\langle P, H\rangle$ is {\em saturated} if each session
  name $s$ occurring in $P$ has a corresponding queue $s:h$ in $H$.
\end{definition}

 We are now ready for defining our $\calL$-bisimulation, expressing
indistinguishability by an $\calL$-observer. {Unlike early
  definitions of $\calL$-bisimulation, which only allowed the ``high
  state'' to be changed at the start of computation, our definition allows
  it to be changed at each step, to account for dynamic contexts~\cite{csfw02}.}

\begin{definition}{{\bf ($\calL$-Bisimulation)}}\text{~}\\
\label{def:low-bisP}
A symmetric relation ${\calR} \subseteq (\proc\times \proc) $ is a
{\em $\calL$-bisimulation} if $P_1 \,{\calR}\, P_2$ implies, for any 
pair of {monotone} $\,\Qu$-sets $H_1$ and $H_2$ {such that $H_1\loweq H_2$}
{and each $\config{P_i}{H_i}$ is} {saturated}:
\[ \begin{array}{l}\mbox{If
      ${\:\:\:\config{P_1}{H_1}} \redsym
      \nuc{\tilde{\r}}{\config{P'_1}{H'_1}}$, 
    then there exist $P'_2, H'_2$ such~that}\\[3pt]
   \mbox{\hspace{12pt}$\config{P_2}{H_2} \redsym^*{\equiv}\:\:
      \nuc{\tilde{\r}}\config{P'_2}{H'_2}\,$, where $H'_1\loweq
      H'_2\,$ {and} $\,P'_1\,\calR\: P'_2$}.\end{array} \]  
Processes $P_1, P_2$ are {\em $\calL$-bisimilar}, $P_1\lowequiv
P_2$, if $P_1 \:{\calR}\: P_2$ for some $\calL$-bisimulation \calR.
\end{definition}




\noindent
Note that $\tilde{\r}$ may either be the empty string or a single name, 
since it appears after a one-step transition. If it is a name, 
it may either be a service name $a$
(communication 
of a private service) or a fresh session
name $s$ (opening of a new session). {In the latter case},
$s$ cannot occur in $P_2$ and $H_2$ by Barendregt convention.
\acapo
  Intuitively, a transition that adds or removes an \calL-message 
  must be simulated in one or more steps, producing the same
  effect on the $\Qu$-set, whereas a transition that {does not affect
${\calL}$-messages} may be simulated by inaction. {In such case,
the structural equivalence $\equiv$ may be needed in case the first process has
created a restriction.}
The notions of $\calL$-security and security are now defined in the standard way:

\begin{definition}{{\bf (Security)}} \label{def:low-prog-secP}\label{def:secP}
\begin{myenumerate}
\item A {process}  is {\em $\calL$-secure} if it is
  $\calL$-bisimilar with itself.
\item A {process} is {\em secure} if it is $\calL$-secure for
every $\calL$.
\end{myenumerate}
\end{definition}

\noindent
The need for considering all downward-closed sets $\calL$
is justified by the following example.
\begin{myexample}
\label{example:need-all-LP}
Let
  $\calS=\set{\bot,\lev,\top}$ where $\bot\sleq\lev\sleq\top$
  and 
  
    \processes{\begin{array}{lll}
P&=&\nsi{\Ia}{2}\pc\sa\Ia1{\Ias_1}{P_1}\pc\sa\Ia2{\Ias_2}{P_2}\\
  P_1&=&\inpDec{\Ias_1}{2}{\x^{\top}}\ifthenelse{\x^{\top}}{\outDec{\Ias_1}{2}{\false^\lev}{\inact}}{\outDec{\Ias_1}{2}{\true^\lev}{\inact}}\\
  P_2&=&\outDec{\Ias_2}{1}{\true^\top}{\inact}
  \end{array}}
  \noindent
  The process $P$ is $\set\bot$-secure 
and $\calS$-secure, 
but it is
  not $\set{\bot,\lev}$-secure, since there is a flow from level
  $\top$ to level $\lev$ in $P_1$, which is detectable by
a $\set{\bot,\lev}$-observer but not by a
$\set{\bot}$-observer. We let the reader verify this fact formally,
possibly after looking at the next example.
\end{myexample}

We show next that
an input of level $\ell$ should not be followed by an action of level
$\ell'\not \geq\ell$:
\begin{myexample}{\em (Insecurity of high input followed by low action)}\\
\label{ex:high-low}
Consider the process $P$ and the $\Qu$-sets $H_1$ and $H_2$,
where $H_1 =_{\set\bot} H_2$:

\processes{\begin{array}{l}
P = \inp{s[2]}{x^\top}{1}{ \outs{s[2]}{\true^\bot}{1}  {\inact}},\quad
 H_1=\set{\s:(1,2,\true^\top)} \qquad
 H_2 = \set{\s:\varepsilon}
\end{array}}

\noindent
Here we have $\red{\config{P}{H_1}}{\config{  \outs{s[2]}{\true^\bot}{1}
    {\inact}}{ \set{\s:\varepsilon}}} = \config{P_1}{H'_1}$, 
while $\config{P}{H_2}\,\,\not\!\!\longrightarrow$. Since
$H'_1 = \set{\s:\varepsilon} = H_2$, we can proceed
with $P_1 = \outs{s[2]}{\true^\bot}{1} {\inact}$ and $P_2 = P$. 
Take now
$K_1 = K_2 = \set{\s:\varepsilon}$. Then
$\red{\config{P_1}{K_1}}{\config{\inact}{
    \set{\s:(2,1,\true^\bot)}}}$, 
while  $\config{P_2}{K_2}\,\,\not\!\!\longrightarrow$. Since
$K'_1 =   \set{\s:(2,1,\true^\bot)}\not=_{\set\bot}
\set{\s:\varepsilon} = K_2$, 
$P$ is not ${\set\bot}$-secure.
\acapo
With a similar argument we may show that $Q= \inp{s[2]}{x^\top}{1}{
\inp{s[2]}{\y^\bot}{1} {\inact}}$ is not ${\set\bot}$-secure.
\end{myexample}

The need for security levels on value variables are justified by the following example.

\begin{myexample}{\em (Need for levels on value variables)}\\
\label{example:queue-condition}
Suppose we had no levels on value variables. Consider the process,
  which should be secure:

\processes{
P \ =\
\inpDec{s[1]}{2}{\x}{\inpDec{s[1]}{2}{\y}{\inact}} 
\: \pc\: 
\outDec{s[2]}{1}{\true^\bot}{\outDec{s[2]}{1}{\true^\bot}{\inact}}
}
\noindent
Let
$ H_1 = \{s: (2,1, \true^\top)\} \loweq \{s: \varepsilon\} = H_2$.
Then the transition:

\processes{\config{P}{H_1} \redsym
\,\config{{\inpDec{s[1]}{2}{\y}{\inact} \: \pc\: 
\outDec{s[2]}{1}{\true^\bot}{\outDec{s[2]}{1}{\true^\bot}{\inact}}}  }
{\{s: \varepsilon\}} = \config{P_1}{H'_1}}
\noindent
could not be matched by $\config{P}{H_2}$.
In fact, the first component of $P$ cannot move in $H_2$, and each
computation of the second component yields an $\calL$-observable 
$H'_2$ such that $H'_1 \not=_{\set\bot} H'_2$. Moreover, $P$ cannot
stay idle in $H_2$, since $P$ is not $\calL$-bisimilar to $P_1$ (as it
is easy to see by a similar reasoning). 
{By adding the level $\bot$ to the variables $\x$ and $\y$}, {we
  force the second component to move first} {in both \config{P}{H_1} and \config{P}{H_2}.}
\end{myexample}

Interestingly, an insecure component may be
  ``sanitised'' by its context, so that the insecurity is not
  detectable in the overall process. Clearly, in case of a deadlocking
  context, the insecurity is masked simply because the dangerous part
  is not executed. However, the curing context could also be
  a partner of the insecure component, as shown by the next example.
This example is significant 
because it constitutes a non trivial case of a
process that is secure but {\em not safe}, as will be further
discussed in Section~\ref{MonSemSaf}.

 \begin{myexample}{\em (Insecurity sanitised by parallel context)}\\
\label{ex:high-lowS}
Let $R$ be obtained by composing the process $P$ of Example~\ref{ex:high-low}
in parallel with a dual process $\overline{P}$, and consider again the $\Qu$-sets $H_1$ and $H_2$,
where $H_1 =_{\set\bot} H_2$:

\processes{\begin{array}{c}
R = \,P\pc \overline{P}\, =\, \inp{s[2]}{x^\top}{1}{ \outs{s[2]}{\true^\bot}{1}  {\inact}} \:\pc\:
\outs{s[1]}{\true^\top}{2}{ \inp{s[1]}{\y^\bot}{2}  {\inact}}
\\
 H_1=\set{\s:(1,2,\true^\top)} \qquad \qquad
  H_2 = \set{\s:\varepsilon}
\end{array}}
\noindent
Then the move  $\red{\config{P\pc \overline{P}}{H_1}}{\config{
    \outs{s[2]}{\true^\bot}{1} {\inact}\pc \overline{P}}{ \set{\s:\varepsilon}}}$
can be simulated by the sequence of two moves

\processes{\begin{array}{l}\red{\config{P\pc \overline{P}}{H_2}}{\config{P\,\pc\,
    \inp{s[1]}{\y^\bot}{2}{\inact}}{ \set{\s:(1,2,\true^\top)} }}\\
    \redsym {\config{\outs{s[2]}{\true^\bot}{1}{\inact}\pc
    \inp{s[1]}{\y^\bot}{2}{\inact}  }{ \set{\s:\varepsilon}}},\end{array}}
    \noindent
where $H'_1 = H'_2 = \set{\s:\varepsilon}$.\\
Let us now compare the processes
$R_1 =  \outs{s[2]}{\true^\bot}{1} {\inact}\pc \overline{P}\,$ 
and $\,R_2 = \outs{s[2]}{\true^\bot}{1}{\inact}\pc
    \inp{s[1]}{\y^\bot}{2}{\inact} $.\\ 
    Let $K_1, K_2$ be monotone $\Qu$-sets containing a queue $s:h$ and
    such that $K_1 =_{\set\bot} K_2$. Now, if $\config{R_1}{K_1}$
    moves first, either it does the high output of $\overline{P}$, in
    which case $\config{R_2}{K_2}$ replies by staying idle, since the
    resulting processes will be equal and the resulting queues $K'_1,
    K'_2$ will be such that $K'_1 =_{\set\bot} K'_2$, or it executes its first
    component, in which case $\config{R_2}{K_2}$ does exactly the
    same, clearly preserving the $\set\bot$-equality of $\Qu$-sets,
    and it remains to prove that $\overline{P} =
    \outs{s[1]}{\true^\top}{2}{ \inp{s[1]}{\y^\bot}{2} {\inact}}$ is
    $\bot$-bisimilar to $\inp{s[1]}{\y^\bot}{2} {\inact}$.  But this
    is easy to see since if the first process moves, the second may
    stay idle, while if the second moves, the first may simulate it in
    two steps.
\acapo
Conversely, if $\config{R_2}{K_2}$ moves first, either it executes
its second component (if the queue allows it), in which case $\config{R_1}{K_1}$ simulates it
in two steps, or it executes its first component, in which case we
are reduced once again to prove that $\overline{P} =
\outs{s[1]}{\true^\top}{2}{ \inp{s[1]}{\y^\bot}{2}  {\inact}}$ is
  $\bot$-bisimilar  to $\inp{s[1]}{\y^\bot}{2}  {\inact}$.
\end{myexample}

\mysection{Monitored Semantics}
\label{monsem}
\begin{mytable}
{\small \centering
\begin{tabular}{l}
$\Mterm{\mlev_1}{\sa {a} {1}{\alpha_1}{\PP_1}}
        \pc...\pc\Mterm{\mlev_n}{\sa{a} {n}{\alpha_n}{\PP_n}} \pc \Mterm{\mlev_{n+1}}{\nsi{a}{n}}
      \Mredsym$\\ 
      \hfill $(\nu {\swl})\config{\Mterm{\mlev}{
        \PP_{1}\sub{\si{\s} {1}}{\alpha_{1}}} \pc ...\pc
        \Mterm{\mlev}{\PP_n\sub{\si{\s}{n}}{\alpha_n}}}{ \sS : \eq}$  \\
        \hfill where $\mlev=\bigsqcup_{i\in \{1 \ldots n+1\}} \mlev_i$
        \hfill [MLink]\\
\\[0.8mm]
         \rederr{\mlev \sleq \lev}{\config{\Mterm{\mlev}{\outDec{s[\p]}{\pset}{\e}{\PP}}}{ \mqueue{\sSp}{\queue}}}
    {\config{\Mterm{\mlev}{\PP }}{\mqueue{\sSp}{\qtail{\valheap{\dle{v}{\lev}{\lev'} }{\p}{\p}} }} }\\
\qquad  where $\e\downarrow v^{\lev} $     
                \hfill  [MSendV] 
\\[0.8mm]
    \rederr{\mlev \sleq \lev}{\config{\Mterm{\mlev}{\inpDec{\s[\q]}{\p}{\x^{\lev}}{\PP} }}{\qpopp{\valheapq{\dle{v}{\lev}{\lev}}{\p}}}}
    { \config{\Mterm{\lev}{\PP\subst{v}{\x} }} {
    \mqueue{\sSp}{\queue}}}
     \\   \hfill [MRecV]
\\[0.8mm]
 \rederr{\mlev \sleq \lev}{\config{\Mterm{\mlev}{\outDecl{s[\p]}{\pset}{\Ia}{\PP}}}{ \mqueue{\sSp}{\queue}}}
    {\config{\Mterm{\mlev}{\PP }}{\mqueue{\sSp}{\qtail{\valheap{\Ia^\lev }{\p}{\p}} }} }     
           \\    \hfill  [MSendS] 
\\[0.8mm]
   \rederr{\mlev \sleq \lev}{\config{\Mterm{\mlev}{\inpl{\s[\q]}{\vs}{\p}{\PP} }}{\qpopp{\valheapq{\Ia^\lev}{\p}}}}
    { \config{\Mterm{\lev}{\PP\subst{\Ia}{\vs} }} {
    \mqueue{\sSp}{\queue}}}
    \\    \hfill [MRecS]
\\[0.8mm]
    \rederr{\mlev \sleq \lev}{ \config{\Mterm{\mlev}{\sdlpr{\s[p]}{\s'[\p']}{\q}{\PP} }}{ \mqueue{\sSp}{\queue}}}
    {{\config{\Mterm{\mlev}{\PP} }{  \mqueue{\sSp}{\qtail{\delheap{\si{\s'}{\p'}^{\lev}}{\q}{\p}}} }}}  
               \\ \hfill [MSendC]
\\[0.8mm]
    \rederr{\mlev \sleq \lev}{\config{\Mterm{\mlev}{ \rdl{\s[\q]}{\Ias
            }{\p}{\PP}}}{ \qpopp{\delheap{\si{\s'}{\p'}^{\lev}}{\q}{\p}}}}
   {\config{\Mterm{\lev}{\PP\subst{\si{\s'}{\p'}}{\Ias}} }{ \stdqueuep}}\\       \hfill  [MRecC]
\\[0.8mm]
    \rederr{\mlev \sleq \lev}{\config{\Mterm{\mlev}{\lselp{\s[\p]}{\lambda}{\PP}}}{ \stdqueuep}}
   {\config{\Mterm{\mlev}{\PP }}{\qappendp{\labheap{\lambda^{\lev}}{\p}{\p}}  }}\\
        \hfill  [MLabel]
\\[0.8mm] 
 \rederr{\mlev \sleq \lev}{ \config{
     \Mterm{\mlev}{\lambdabranchp{\s[\q]}{\p} } }
   {\qpopp{\valheapq{\lambda_{i_0}^{\lev}}{\p}}} }
{ \config{ \Mterm{\lev}{\PP_{i_0
}} }{ \mqueue{\sSp}{\queue}} }\\ \qquad  \text{where  $i_0 \in I$}
        \hfill  [MBranch]\\[0.8mm]
  \Mred{\Mterm{\mlev}{\ifthenelse{\e}{\PP}{\Q}}}{\Mterm{\mlev \join \lev}{\PP}}
 \qquad \mbox{~~~~if $\e \downarrow \true^{\lev}$}
\\
\Mred{\Mterm{\mlev}{\ifthenelse{\e}{\PP}{\Q}}}{\Mterm{\mlev \join \lev}{\Q}}
\qquad \mbox{~~~~if $\e \downarrow \false^{\lev}$}
        \hfill [MIf-T, MIf-F]
\\[1.5mm]
\Mred{\Mterm{\mlev}{(\DdefDlp (\proccallw{X}{\e}{\si{\s}{\p}} ))}}
   {\DdefDlp
    (\Mterm{\mlev}{\PP\subst{\at{\vap}}{\at{\x}}\subst{\atw{\si{\s}{\p}}}{\alpha})}}\\
    \qquad where
    $\ptilde{\e}\downarrow \at{\va}$ 
           \hfill [MDef]\\
\\[0.8mm]
\Mredbin{\config{\Mp}{\hH}}{\nuc{\tilde{\swl}}\config{\Mp'}{\hH'}}
\Implies\\
\Mredbin{\config{\defD(\Mp\pc\Mp'')}{\hH}}
{\nuc{\tilde{\swl}}\config{\defD(\Mp'\pc\Mp'')}{\hH'}}\\
        \hfill [MDefin]
\\[1mm]
\Mredbin{C}{\nuc{\tilde{\swl}}{C'}} and $\neg~C''\error$
\Implies\Mredbin{\nuc{\tilde{\r}}(\,C\,\parcon\: C'')}{\nuc{\tilde{\r}}{\nuc{\tilde{\swl}}(\,C'\,\parcon\: C'')}}\\
$C\error\Implies\nuc{\tilde{\r}}(\,C\,\parcon\: C')\error$
        \hfill  [MScopC]\\[2pt]       
\end{tabular}
\caption{Monitored reduction rules.}\label{tab:Mreduction}}
\bigskip
\end{mytable}
In this section we introduce the monitored semantics for our
calculus. This semantics is defined on {\em monitored processes} $M, M'$, whose
syntax is the following, assuming $\mu\in\calS$:

\processes{ M::= \Mterm{\mlev} \PP \separ \Mp\pc\Mp \separ
\nuc{\Ia}\Mp \separ \defD\Mp
}
\noindent
In a monitored process $\Mterm{\mlev}{P}$, 
the level $\mu$ that tags $P$ is called 
the {\em monitoring level} for $P$. It
controls the execution of $P$ by blocking any communication of
level $\ell\not\geq\mu$. Intuitively,
$\Mterm{\mlev}{P}$ represents a partially executed process, and $\mu$
is the join of the levels of received objects (values, labels or
channels) and of tested conditions up
to this point in execution. 
\acapo
The monitored semantics is defined on monitored configurations
$C= \config{\Mp}{\hH}$. By abuse of notation, we use 
the same symbol $C$ for standard and monitored configurations.
\acapo
The semantic rules define simultaneously a reduction relation
$\Mredbin{C}{C'}$
and an error predicate $C\,\dagger$ on monitored configurations. 
As usual, the semantic rules are applied modulo a {\em structural
  equivalence} $\equiv$.
The new specific structural rules are:

\processes{
\begin{array}{c}
\Mterm{\mlev}{(\PP_1\pc \PP_2)} \,\equiv\,
\Mterm{\mlev}{\PP_1}\pc\Mterm{\mlev}{\PP_2} \qquad \qquad
 C\dagger  \ \ \wedge\ \ C\equiv C'\ \ \implies\ \ C'\dagger
\end{array}
}
The reduction rules of the monitored semantics are given in
Table~\ref{tab:Mreduction}. 
Intuitively, the monitoring level is initially $\bot$ and gets
increased each time a test of higher or incomparable level or an input of higher level
is crossed. Moreover, if $\config{\Mterm{\mu}{P}}{H}$ attempts to 
perform a communication action of level 
$\ell\not\geq\mu$,
 then $\config{\Mterm{\mu}{P}}{H}\, \dagger$.
We say in this case that the reduction produces error. 
\acapo
The reason why the monitoring level should take into account the
  level of inputs is that, as argued in Section~\ref{MonSemSec}, the
  process
  $\inpDec{s[1]}{2}{\x^\top}{\outDec{s[1]}{2}{\true^\bot}{\inact}}$ is
  not secure. Hence it should not be safe either.\acapo
One may wonder whether monitored processes of the form 
$\Mterm{\mlev_1}{\PP_1}\pc\Mterm{\mlev_2}{\PP_2}$, where $\mu_1
\neq\mu_2$, are really needed.
The following example shows that, in the presence of concurrency, a
single monitoring level (as used for instance
in~\cite{Boudol08}) would not be enough.

\begin{myexample}{\em (Need for multiple monitoring levels)}\\
  Suppose we could only use a single monitoring level 
    for the parallel process $P$ below, which should intuitively be safe. Then a computation of
    $\Mterm{\bot}{P}$ would be successful or not depending on the
    order of execution of its parallel components:

\processes{\begin{array}{l}
P_1=\outDec{\Ias_1}{2}{\true^\bot}{\inact} \qquad
P_2=\inpDec{\Ias_2}{1}{\x^{\bot}}{\inact} \\
P_3=\outDec{\Ias_3}{4}{\true^\top}{\inact}\qquad 
P_4=\inpDec{\Ias_4}{3}{\y^{\top}}{\inact}\\
P=\nsi{\Ia}{4}\pc\sa\Ia1{\Ias_1}{P_1}\pc\sa\Ia2{\Ias_2}{P_2}\pc\sa\Ia3{\Ias_3}{P_3}\pc\sa\Ia4{\Ias_4}{P_4}
\end{array}}
\noindent
Here, if $P_1$ and $P_2$ communicate first, we would have the
successful computation:

\processes{ \Mterm{\bot}{P}\Mredsymstar(\nu\s)\config{\Mterm{\bot}
{(P_3\sub{\si{\s} {3}}{\alpha_{3}} \pc P_4\sub{\si{\s}{4}}{\alpha_{4}})}}{\s:\epsilon}\Mredsym(\nu \s)\config{\Mterm{\top}{\inact}}{\s:\epsilon}}
\noindent
Instead, if $P_3$ and $P_4$ communicate first, then we would run into error:

\processes{ \Mterm{\bot}{P}\Mredsymstar(\nu\s)\config{\Mterm{\top}
{(P_1\sub{\si{\s} {1}}{\alpha_{1}} \pc
  P_2\sub{\si{\s}{2}}{\alpha_{2}})}}{\s:\epsilon}
\error}
\noindent
{Intuitively, the monitoring level resulting from the
  communication of $P_3$ and $P_4$ should not constrain the
  communication of  $P_1$ and $P_2$, since there is no causal
  dependency  between them.}
{\noindent Allowing}  {different monitoring levels for
  different parallel components}, {when $P_3$ and $P_4$ communicate first we get:

\processes{ \Mterm{\bot}{P}\Mredsymstar(\nu\s)\config{\Mterm{\top}{\inact}\pc\Mterm{\bot}
{(P_1\sub{\si{\s} {1}}{\alpha_{1}} \pc
  P_2\sub{\si{\s}{2}}{\alpha_{2}})}}{\s:\epsilon}
\Mredsymstar(\nu\s)\config{\Mterm{\top}{\inact}\pc\Mterm{\bot}{\inact}}{\s:\epsilon} }
}
\end{myexample}

The following example justifies the use of the join in rule
[MLink]. Session initiation is the only synchronisation operation of
our calculus. Since this synchronisation requires the 
initiator $\nsi{\Ia}{n}$ as well as a complete set of ``service
callers'' $\sa a{\p}{\alpha_1}{\PP_\p},
1\leq\p\leq n$, the monitoring level of each
of them contributes to the monitoring level of the session.
Note that the fact that this monitoring level may be computed
  dynamically as the join of the monitoring levels of the participants exempts us 
  from statically annotating services with levels, as it
  was necessary to do in~\cite{CCDR10} in order to type
the various participants consistently.
\acapo
Consider the process:

\processes{\inp{s[2]}{x^\top}{1}{\ifthenelse{x^\top}{\nsi{\Ib}{2}}{\inact}}\pc\sa\Ib1{\Ibs_1}{\outDec{\Ibs_1}{2}{\true^\bot}{\inact}} \pc \sa\Ib2{\Ibs_2}{\inpDec{\Ibs_2}{1}{\y^{\bot}}{\inact}}}
Here the monitoring level of the conditional becomes $\top$
  after the test, and thus, assuming the {\tt if}
  branch is taken, rule [MLink] will set the monitoring
  level of the session to $\top$. This will block the exchange of the
  $\bot$-value between the last two components.

\begin{myexample}{\em (Need for security levels on transmitted service
    names)}\\
\label{service-level}
\noindent
This example shows the need for security levels on service names in
rules [MSendS] and [MRecS].
  
\processes{
\begin{array}{ll}
\inp{s[2]}{x^\top}{1}
{\ifthenelse{x^\top}{\outDecl{s[2]}{3}{a}{\inact}}
{\outDecl{s[2]}{3}{b}{\inact}}}
\\
\pc\ \inpl{s[3]}{\vs}2{\nsi{\vs}{2}}{}
\\
\pc\ \sa\Ia 1 {\Ias_1}{\outs{\Ias_1}{\true^\bot}2{\inact}} 
\pc\ \sa\Ia 2 {\Ias_2}{\inp{\Ias_2}{\y^\bot}{1}{\inact}}\\
\pc\ \sa\Ib 1 {\Ibs_1}{\outs{\Ibs_1}{\false^\bot}2{\inact}} 
\pc\ \sa\Ib 2 {\Ibs_2}{\inp{\Ibs_2}{\y^\bot}{1}{\inact}}
\end{array}
}
\noindent
This process is insecure because, depending on the high value received
for $x^\top$, it will initiate a session on service $a$ or on
  service $b$, which both
perform a low value exchange.
If $\lev\not=\top$, the monitored semantics will yield error in the outputs
of the first line, otherwise it yields error in the
outputs of the last two lines.
\end{myexample}

{Similar} examples show the need for security levels on
  transmitted channels and labels.
 
\mysection{Safety}
\label{MonSemSaf}
We define now the property of {\em safety}\/ for
monitored processes, from which we derive also a property of 
  safety for processes. We then prove that if a process is safe,
it is also secure.
\acapo
A monitored process may be  ``relaxed'' to an
ordinary process by removing all its monitoring levels.

\begin{definition}{\bf (Demonitoring)}
If $\Mp$ is a monitored process,  its {\em demonitoring}
$\el\Mp$ is defined by:

\processes{\begin{array}{lllclll}
\el{\Mterm{\mlev} \PP}&=&\PP&\qquad&
\el{\Mp_1\pc\Mp_2}&=&\el{\Mp_1}\pc\el{\Mp_2}\\
\el{\nuc{\Ia}\Mp}&=&\nuc{\Ia}\el{\Mp}&&
\el{\defD\Mp}&=&\defD\el{\Mp}
\end{array}}\end{definition} 
Intuitively, a monitored process $\Mp$ is safe if it can mimic at each
  step the transitions of the process $\el\Mp$.
  
\begin{definition}{\bf (Monitored process safety)}\\
\label{mlsafetyP}
\noindent
The safety predicate on monitored
processes is coinductively defined by:\\
\indent $\Mp$ is safe if for any 
monotone
  $\,\Qu$-set $H$ such that
 ${\config{\el\Mp}{H}}$
is saturated:

\processes{\begin{array}{l}\mbox{If
      ${\config{\el\Mp}{H}} \redsym \nuc{\tilde{\r}}{\config{P}{H'}}$}\\[3pt]
    \mbox{then $\config{\Mp}{H}\Mredsym\nuc{\tilde{\r}}{\config{\Mp'}{H'}}$, where $\el{\Mp'} = P$ and  $\,\Mp'$ is
  safe.}\vspace{-2mm}\end{array}}
\end{definition}
\smallskip

\begin{definition}{\bf (Process safety)}
\label{psafetylP}
 A process $P$ is safe\  if $\Mterm{\bot}{P}$ is safe.
\end{definition}

We show now that if a process is safe, then none of its monitored computations starting
 with monitor $\bot$ gives rise to error.
This result
rests on the observation that
$\config{\Mp}{H} \Mredsym$ if and only if $\config{\el{M}}{H}\redsym$
and $\neg\, \config{\Mp}{H}\dagger$, and that if $M$ is safe,
then if a standard communication rule is applicable to
$\el{M}$, the corresponding monitored communication rule is
applicable to $\Mp$.
\medskip

\begin{myproposition}{\bf Safety implies absence of run-time errors}\text{~}\\
\label{no-error}
If $P$ is safe, then
every monitored computation:

\processes{\config{\Mterm{\bot}{P}}{\emptyset} \, = \config{\Mp_0}{H_0}
\Mredsym\nuc{\tilde{\r_1}}{\config{\Mp_1}{H_1}}\Mredsym \cdots 
\nuc{\tilde{\r_k}}{\config{\Mp_k}{H_k}}}
\noindent
is such that $\neg\, \config{\Mp_k}{H_k}\dagger$.
\end{myproposition}

\noindent
Note that the converse of Proposition~\ref{no-error} does not
hold, as shown by the next example. This means that we could not use
absence of run-time errors as a definition of safety, since that would
not be strong enough to guarantee our security property, which allows
the pair of $\calL$-equal $\Qu$-sets to be refreshed at each step (while maintaining $\calL$-equality).

\begin{myexample}
\label{example:sibling}
$\;$\acapo
\processes{\begin{array}{lll}
P&=&\nsi{\Ia}{2}\pc\sa\Ia1{\Ias_1}{P_1}\pc\sa\Ia2{\Ias_2}{P_2}\\
P_1&=&\outDec{\Ias_1}{2}{\true^\top}{\inpDec{\Ias_1}{2}{\x^\top}\inact}\\
P_2&=&\inpDec{\Ias_2}{1}{\z^\top}{\ifthenelse{\z^\top}{\outDec{\Ias_2}{1}{\false^\top}{\inact}}{\outDec{\Ias_2}{1}{\true^\bot}{\inact}}}
\end{array}}
%
Note first that this process is not $\bot$-secure because after $P_1$
has put the value $\true^\top$ in the $\Qu$-set, this value may be
changed to $\false^\top$ while preserving $\calL$-equality of
$\Qu$-sets, thus allowing the {\tt else} branch of $P_2$ to be
explored by the bisimulation.
This process is not safe either,
because our definition of safety mimics
$\calL$-bisimulation by refreshing the $\Qu$-set at each
step. On the other hand, a simple monitored
execution of $\config{\Mterm{\bot}{P}}{\emptyset}$, which uses at each
step  the $\Qu$-set produced at the previous step, would never take the
{\tt else} branch and would therefore always succeed. Hence the simple absence of
run-time errors would not be sufficient to enforce security.
\end{myexample}


In order to prove that safety
implies security, we need some preliminary results.
\begin{lemma}{\bf(Monotonicity of monitoring)}
\label{monot-monit}
Monitoring levels may only increase along execution:
 If $\langle \Mterm{\mlev}{P}, H\rangle \Mredsym \nuc{\tilde{\r}}{\langle
    \Mterm{\mlev'}{P'}{\pc\Mp}, H'\rangle}$, then $\mlev\leq\mlev'$.
 \end{lemma}
 %
\smallskip

{As usual, $\calL$-high processes modify \Qu-sets} {in a way which is transparent} {for $\calL$-observers.}

\begin{definition}[$\calL$-highness of processes]
\label{semh}
A process $P$ is {\em $\calL$-high} if for any monotone $\,\Qu$-set $H$ such
that ${\config{P}{H}}$ is saturated, it satisfies the property:

\processes {\mbox{If $\config{P}{H} \redsym \nuc{\tilde{\r}}{\config{P'}{H'}}$, then $H\loweq H'$ and $P'$ is $\calL$-high.}}

\end{definition}

\begin{lemma}
\label{key}
 If $\Mterm{\mlev}{P}$ is safe and $\mlev\not\in\calL$, then 
$P$ is $\calL$-high.
\end{lemma}

We next define the bisimulation relation that will be used in the
proof of soundness. {Roughly,}
{all monitored processes with a high monitoring level are
  related,} 
  {while the other
  processes} {are related if they are
    congruent.}

\begin{definition}{\bf(Bisimulation for soundness proof: monitored
    processes)}\label{bsmp}\\
\noindent
  Given a downward-closed set of security levels $\mathcal{L}\subseteq
  \calS$, the relation $\RLG$ on monitored processes is defined
  inductively as follows: \smallskip

\noindent
$\Mp_1\,\RLG\,\Mp_2$ if $\,\Mp_1$ and $\Mp_2$ are safe and one of the following holds 
\begin{myenumerate}
\item\label{c1}
$\Mp_1=\Mterm{\mlev_1}{\PP_1}$, $\Mp_2=\Mterm{\mlev_2}{\PP_2}$ and  $\mlev_1,\mlev_2\not\in\calL$; 
\item\label{c2} $\Mp_1=\Mp_2=\Mterm{\mlev}{\PP}$ and $\mlev\in\calL$;\\[-9pt]
\item\label{c3} $\Mp_i = \prod_{j=1}^{m}\Np_j^{(i)}$, where
  $\forall j \in \{1, \ldots, m\}$, $\:\Np_j^{(1)} \,\RLG\,\Np_j^{(2)}$
follows from (\ref{c1}) or (\ref{c2});
  \item\label{c4} $\Mp_i =  \nuc{\Ia}\Np_i$, where
  $\Np_1 \,\RLG\,\Np_2$;
   \item\label{c5} $\Mp_i =  \defD{\Np_i}$, where
  $\Np_1 \,\RLG\,\Np_2$.
  \end{myenumerate}
\end{definition}

\begin{definition}{\bf(Bisimulation for soundness proof:
    processes)}\label{bsp}\\
Given a downward-closed set of security levels $\mathcal{L}\subseteq \calS$, the relation
$\RL$ on processes is defined by:\smallskip

$P_1\RL P_2$ if there are $\Mp_1,\Mp_2$ such that $P_i \equiv\el{\Mp_i}\,$
for $i=1,2$ and $\Mp_1\,\RLG\,\Mp_2$.
\end{definition}

We may now state our main result, namely that safety implies
security. The proof consists in showing that safety implies
\lsecurity, for any $\calL$.  The informal argument goes as
follows. Let ``low'' mean ``in \calL'' and ``high'' mean ``not in
\calL''.  If $P$ is not \lsecure, this means that there are two
different observable low behaviours after a high input or in the two
branches of a high conditional. This implies that there is some
observable low action after the high input, or in at least one of the
branches of the high conditional. But in this case the monitored
semantics will yield error, since it does so as soon as it meets an
action of level $\ell\not\geq\mu$, where $\mu$ is the monitoring level
of the executing component (which will have been set to high after crossing
the high input or the high condition). 

\begin{theorem}{\bf (Safety\ implies security)}\labelx{thm:safety-security} \ If $\PP$ is safe, $P$ is
    also secure.
\end{theorem}

The converse of Theorem \ref{thm:safety-security} does not hold,
  as shown by the process $R$ of Example \ref{ex:high-lowS}.
A more classical example is
$\inp{s[1]}{x^\top}{2}{\ifthenelse{\x^{\top}}{\outDec{s[1]}{2}{\true^\bot}{\inact}}{\outDec{s[1]}{2}{\true^\bot}{\inact}}}$.

\mysection{Conclusion}
\label{conclusion}
{There is a wide literature on the use of
monitors (frequently in combination with types) for assuring security,
but most of this work 
has focussed so far on sequential computations, see for instance \cite{GBJD06,Boudol08,SR09}.}
More specifically, \cite{GBJD06} considers an automaton-based monitoring
mechanism for information flow, combining static and dynamic analyses,
for a sequential imperative while-language with outputs.
The paper \cite{Boudol08}, which provided the initial inspiration for our work,
deals with an ML-like language and uses a single monitoring level
to control sequential executions.
The work \cite{AskSab} shows how to enforce {information-release
policies, which may be viewed as relaxations of noninterference, by a
combination of monitoring and static analysis, in a sequential
language with dynamic code evaluation.
Dynamic security policies and means for expressing them via security
labels have been studied for instance in~\cite{Myers1,Myers2}.
\acapo
In session calculi, concurrency is present not only among participants in a given
session, but also among different sessions running in parallel and
possibly involving some common partners. {Hence, different
monitoring levels are needed to control
different parallel components, and these levels must be joined when
the components convene to start a new session. As we use a
general lattice of security levels (rather than a two level
lattice as it is often done), it may happen that while all the
participants monitors are ``low'', their join is ``high'',
constraining all their exchanges in the session to be high too.
Furthermore, we deal with structured memories
(the $\Qu$-sets).  In this sense, our setting is slightly more complex
than some of the previously studied ones.}  Moreover, a peculiarity of
session calculi is that data with different security levels are
transmitted on the same channel\footnote{Each session channel is used
``polymorphically'' to send objects of different types and levels,
since it is the only means for a participant to communicate with the
others in a given session.}  (which is also the reason why security
levels are assigned to data, and not to channels).
Hence, although the intuition behind monitors is
rather simple, its application to our calculus is not completely
straightforward. 
\acapo 
Session types have been proposed for a variety of calculi and
languages. We refer to \cite{DL10} for a survey on the session type
literature. However, the integration of security requirements into
session calculi is still at an early stage. A type system assuring
that services comply with a given data exchange security policy is
presented in \cite{LPT07}. Enforcement of integrity properties in
multiparty sessions, using session types, has been studied
in~\cite{corinsessions09,planulconcur09}. These papers propose a
compiler which, given a multiparty session description, implements
cryptographic protocols that guarantee session execution integrity.
\acapo We expect that a version of our monitored semantics, enriched
with labelled transitions, could turn useful to the programmer, either
to help her localise and repair program insecurities, or to
deliberately program well-controlled security transgressions,
according to some dynamically determined condition.  To illustrate
this point, let us look back at our medical service example of
Figure~\ref{medical-example} in Section~\ref{introex}.  In some
special circumstances, we could wish to allow the user to send her
message in clear, for instance in case of an urgency, when the user
cannot afford to wait for {data encryption and decryption.}
Here, if in the code of \Us\
we replaced the test on
$\mathit{\textsl{gooduse}\,(\textit{form}^\top)}$ by a test on
$\textsl{no-urgency}^\top \wedge
\,\mathit{\textsl{gooduse}\,(\textit{form}^\top)}$, then in case of
urgency we would have a security violation, which however should
not be considered incorrect, given that it is expected by the
programmer.  A labelled transition monitored semantics, whose labels
would represent security errors, would then allow the
programmer to check that her code's insecurities are exactly the
expected ones. {This labelled semantics could also be used to
control error propagation, thus avoiding to block the execution of the
whole process in case of non-critical or limited errors. In this
case, labels could be recorded in the history of the process 
and the execution would be allowed to go on, postponing error
analysis to natural breaking points (like the end of a session).}

\smallskip

\noindent
{\bf Acknowledgments}\\
We would like to thank Kohei Honda, Nobuko Yoshida and
the anonymous referees for helpful feedback.

\bibliographystyle{eptcs}
\bibliography{session-vshort}

\end{document}